\documentclass[letterpaper, 10 pt, conference]{ieeeconf}  % Comment this line out if you need a4paper

\IEEEoverridecommandlockouts                              % This command is only needed if 
\def\BibTeX{{\rm B\kern-.05em{\sc i\kern-.025em b}\kern-.08em
    T\kern-.1667em\lower.7ex\hbox{E}\kern-.125emX}}
 
 \usepackage{cite}
\usepackage{amsmath,amssymb,amsfonts}
\usepackage{empheq}
\usepackage{algorithmic}
\usepackage{graphicx}
\usepackage{textcomp}
\usepackage{xcolor}
\usepackage{tabularx}
\usepackage[font=footnotesize,skip=4pt]{caption}

\title{\LARGE \bf
Persistently Exciting Online Feedback Optimization Controller with Minimal Perturbations*
}
 
\author{Tore Gude$^{1,\dagger}$, Marta Anna Zagorowska$^{2}$ and Lars Struen Imsland$^{1}$% <-this % stops a space
\thanks{*This project is part of BRU21-NTNUs Research and Innovation Program in Digital and Automation Solutions for the Offshore Energy Industry (https://www.ntnu.edu/bru21/). Financial support from Harbour Energy is acknowledged. M. Zagorowska also kindly acknowledges funding from Marie Curie Horizon Postdoctoral Fellowship project RELIC (grant no 101063948).}% <-this % stops a space
\thanks{$^{1}$Department of Engineering Cybernetics, Norwegian University of Science and Technology, Trondheim, Norway,  {\tt\small tore.gude,lars.imsland@ntnu.no}, $^{2}$Delft Center for Systems and Control, TU Delft, Delft, Netherlands, {\tt\small m.a.zagorowska@tudelft.nl}}
\thanks{$^\dagger$Corresponding author}
}

\begin{document}

\maketitle
\thispagestyle{empty}
\pagestyle{empty}

\begin{abstract}
This paper develops a persistently exciting input generating Online Feedback Optimization (OFO) controller that estimates the sensitivity of a process ensuring minimal deviations from the descent direction while converging. This eliminates the need for random perturbations in feedback loop. The proposed controller is formulated as a bilevel optimization program, where a nonconvex full rank constraint is relaxed using linear constraints and penalization. The validation of the method is performed in a simulated scenario where multiple systems share a limited, costly resource for production optimization, simulating an oil and gas resource allocation problem. The method allows for less input perturbations while accurately estimating gradients, allowing faster convergence when the gradients are unknown. In the case study, the proposed method achieved the same profit compared to an OFO controller with random input perturbations, and $1.4\%$ higher profit compared to an OFO controller without input perturbations.

\end{abstract}

\section{Introduction}
Increasingly complex industrial processes must be optimally operated to be competitive and minimize environmental impacts. The availability of sensor data, computational power, and advances in machine learning motivate utilizing real-time measurements to find the optimal operating point of a process even in the presence of model-plant mismatch and external disturbances. To compensate for model-plant mismatch, the measurements must contain enough information about the system, formalized through persistency of excitation \cite{b3}. 

In this work, we propose a method for generating persistently exciting inputs while minimally disturbing an Online Feedback Optimization (OFO) controller. Persistently exciting input signals are necessary for accurate model estimation, which is used by the OFO controller. The proposed controller is tested in a simulated scenario in which multiple systems utilize the same shared, limited, resource for production optimization, reflecting resource allocation challenges in the oil and gas industry.

An OFO controller optimizes a system in a feedback loop \cite{b3}. Using optimization in feedback loops has recently been investigated and has shown promising results for optimal power system operation \cite{b17}. OFO controllers that generate persistently exciting input have also been investigated in \cite{b3} and \cite{b7}. However, these approaches involve random perturbations and do not consider possible adverse effects of these perturbations. 

An overview of other methods that combine optimization with feedback control and their relation to OFO is provided in \cite{b9}, including a comparison of real-time optimization methods such as extremum seeking and modifier adaptation. Furthermore, an overview of extremum seeking approaches including treatment of persistent excitations is given in \cite{Scheinker_2024}.

Our contribution is a persistently exciting input generating OFO controller with minimal perturbations while converging toward the optimal operating point. This approach eliminates the need for random perturbations, which an operator or production engineer may be reluctant to introduce into a critical feedback loop. For the controller to efficiently find a solution, a nonconvex full rank constraint is relaxed using linear constraints and penalization. While this looses the theoretical guarantee for generating persistently exciting inputs, it renders the problem efficient to solve and demonstrates effectiveness in the case study.

The remainder of this paper is organized as follows: In Section \ref{sec:Prelimiaries} we present preliminary theory of OFO and default gradient estimation. In Section \ref{sec: pers ex OFO}, we present our persistently exciting input generating OFO controller, with case study results given in Section \ref{sec: example study}. Finally, the paper is concluded in Section \ref{sec: Conclusion}.

\section{Preliminaries}
\label{sec:Prelimiaries}
\subsection{Online Feedback Optimization}
The recent research in optimization in closed loop is based on three main benefits, 1) optimize systems with inaccurate data and time-varying parameters, 2) obey constraints without an accurate model of the whole system and 3) eliminate the need for exogenous setpoints and reference signals \cite{b9}. OFO is a real-time optimization scheme that solves an optimization problem by iteratively driving a system closer to the optimal operating point. This is done utilizing steady-state measurements in a feedback loop instead of a model of the system. 

Given a general control system with inputs $u \in \mathbb{R}^{n_u}$, outputs $y \in \mathbb{R}^{n_y}$ and a steady-state mapping $y = h(u)$, OFO solves the following optimization problem, 
\begin{subequations} \label{eq: mod 0}
\begin{align}
         \min_{u, y} &\quad \Phi\left(u,y\right) \label{eq:OFOObjective}\\
         \textrm{s.t. } &  u \in \mathcal{U}\label{eq:InputCstr}\\
       & y \in \mathcal{Y}\label{eq:OutputCstr}\\
       & y = h\left(u\right), \label{eq: y=h(u)}
\end{align}
\end{subequations}
where $\Phi:\mathbb{R}^{n_u}\times\mathbb{R}^{n_y}\rightarrow\mathbb{R}$ is a continuously differentiable objective function, $h:\mathbb{R}^{n_u}\rightarrow\mathbb{R}^{n_y}$ is a continuously differentiable nonlinear input-output mapping and $\mathcal{U}$ and $\mathcal{Y}$ are the sets of input and output constraints, respectively. To solve the optimization problem \eqref{eq: mod 0}, OFO typically actuates the real system in the direction of the negative descent of the objective function until a local minimum of \eqref{eq: mod 0} is found. To account for input and output constraints, in this work OFO uses projected gradient descent, which projects the gradient onto the closest direction in the tangent cone, $\mathcal{T}_u$,
\begin{equation}\label{eq: projected gradient descent}
    w_t = \underset{w \in \mathcal{T}_u}{\mathrm{arg\,min}} \lVert w+ \nabla_u\Phi^\top\left(u,y\right) \rVert^2_2,
\end{equation}
 ensuring safe and feasible operation while converging \cite{b4}. Then the solution to \eqref{eq: projected gradient descent}, $w_t$, is applied as an input actuation
\begin{equation}\label{eq: update step w}
    u_{t+1} = u_t + \alpha w_t,
\end{equation}
where $\alpha$ is a constant stepsize and
\begin{equation}\label{eq:DescentAll}
    \nabla_u\Phi^\top\left(u,y\right) = \frac{\partial \Phi^\top\left(u,y\right) }{\partial u} +\nabla_u h(u)^\top\frac{\partial \Phi^\top\left(u,y\right) }{\partial y}
\end{equation}
is the gradient of the objective function with respect to $u$. An analysis of the impact of the step size $\alpha$ in OFO was done by \cite{b4}. When the input-output mapping (\ref{eq: y=h(u)}) is explicitly known, the derivatives in \eqref{eq:DescentAll} can be found analytically. To obtain the descent direction \eqref{eq:DescentAll}  given an unknown input-output mapping $y = h(u)$, the Jacobian, $\nabla_u h$ needs to be estimated \cite{b3}. It can be shown that OFO will converge to an optimal operating point with the estimated $\nabla_u h$ and with disturbances affecting the system \cite{b18}.

\subsection{Gradient Estimation} \label{sec: Kalman}
For estimation of $\nabla_u h$, we use a similar approach as in \cite{b3} where its column-wise vector representation  $\hat{h}_t$, is estimated through a recursive least-squares algorithm,

 \begin{subequations}
    \begin{align}
        K_{t-1} =&
        \Sigma_{t-1} U^\top_{\Delta,{t-1}}\left(\Sigma_{m,t-1} + U_{\Delta,t-1}\Sigma_{t-1} U_{\Delta,t-1}^\top\right)^{-1} \\
        \hat{h}_t =&
        \hat{h}_{t-1} + K_{t-1}\left(\Delta y_{t-1}-U_{\Delta,t-1}\hat{h}_{t-1}\right) \\
        \Sigma_t =&(I_{n_un_y}-K_{t-1}U_{\Delta,t-1})\Sigma_{t-1}(I_{n_un_y} - K_{t-1}U_{\Delta , t-1})^\top \notag \\
        &\quad + K_{t-1}\Sigma_{m,t-1}K_{t-1}^{T} + \Sigma_{p,t-1}, 
    \end{align}
\end{subequations}
 where $\Delta y_{t-1} = y_t - y_{t-1}$, $\Delta u_{t-1} = u_t-u_{t-1}$, $U_{\Delta,t-1} = \Delta u_{t-1} \otimes I_{n_y}$, where $\otimes$ is the Kronecker product, $I_{n_y}$ is the identity matrix of dimension $n_y$,  $\Sigma_{p,t} = \Sigma_{p1} + \Sigma_{p2}\lVert \Delta u_t \rVert_2^2$ is the covariance matrix of the Gaussian process noise and $\Sigma_{m,t} = \Sigma_{m1} + \Sigma_{m2}\lVert \Delta u_t \rVert_2^2 + \Sigma_{m3}\lVert \Delta u_t \rVert_2^4$ is the covariance matrix of the Gaussian measurement noise.
 The parts $\Sigma_{p2}, \Sigma_{m2}$ and $\Sigma_{m3}$ of the covariance matrices are proportional to $\lVert\Delta u_{t}\rVert_2$, so that larger variations in input are associated with greater uncertainty. The covariance matrices also have parts independent of the input, $\Sigma_{p1}$ and $\Sigma_{m1}$ to account for uncontrolled random changes. In this paper, $\Sigma_{m,0}$ and $\Sigma_{m,0}$ are initialized as identity and $\hat{h}_0$ as a vector of ones.

\section{Persistently Exciting OFO Controller}\label{sec: pers ex OFO}
To ensure accurate estimation of $\nabla_u h$, the inputs $\Delta u$ need to be persistently exciting, enabling exploration of a range of outputs $\Delta y$ \cite{b3}.  

\subsection{Random Perturbations}
In the OFO formulation from \cite{b3}, the descent direction \eqref{eq:DescentAll} is perturbed with Gaussian noise, $s_t \sim \mathcal{N}\left(0,\sigma\right)$, to ensure persistency of excitation with high probability. Combining the OFO controller \eqref{eq: projected gradient descent} from \cite{b4} with the perturbation from \cite{b3} the following formulation is an OFO controller that generates persistently exciting input signals with the same high probability as in \cite{b3},
\begin{subequations}\label{eq: orig opt}
                \begin{align}
                w_t =& \underset{w \in \mathbb{R}^{n_u}}{\mathrm{arg\,min}} \lVert w + \nabla_u\Phi^\top\left(u,y\right) + s_t \rVert^2_2 \label{eq: orig opt obj}\\
                % \text{s.t.} \quad \underline{\boldsymbol{u}} \leq \overbrace{\boldsymbol{u}_t +\alpha w_t}^{u_{t+1}}\leq \overline{\boldsymbol{u}}
                 \text{s.t.} &\quad u_t +\alpha w\in \mathcal{U}\\
                &\quad y_t+\alpha \nabla_u h_t w \in \mathcal{Y},
            \end{align}
        \end{subequations}
where $y_t+\alpha \nabla_u h_t w_t$ is the estimated output $y_{t+1}$. Persistently exciting OFO controllers have also been investigated in \cite{b7}. Here, if the input actuation during an iteration of the OFO controller is not persistently exciting, it is arbitrarily perturbed in a direction giving persistently exciting input. This approach is similar to the OFO controller we present in the following subsections, where we also perturb the descent direction only when the candidate solution is not persistently exciting. However, the methods differ in how the perturbation is determined. 

\subsection{Persistently Exciting Input Design}
The persistency of excitation condition \cite{b14} states that $\Delta u_t$ is persistently exciting if there exists a $T>0$, such that for all $t>0$, the matrix formed by columns $\Delta u_{t-i}, \forall i \in \{0, \ldots, T\}$ has full rank, rank equal to the dimension of the input, $n_u$. This means that $\mathbb{R}^T$ can be spanned by $\Delta u_i, \forall i \in \{0, \ldots, T\}$ in $T$ steps when $\Delta u_i, \forall i \in \{0, \ldots, T\}$ is persistently exciting. To have a persistently exciting OFO controller, we want to achieve persistency of excitation with a minimal number of time steps, $T=n_u$, resulting in a square matrix. Ideally, full rank would be achieved following the gradient descent without any perturbation as in (\ref{eq: update step w}). Any nonzero perturbation that is nonparallel to the gradient descent in \eqref{eq:DescentAll} would affect the projection \eqref{eq: orig opt obj}  which could lead to slower convergence. In particular, we want to avoid perturbing the gradient descent in a given direction if it already gives a persistently exciting input. We therefore reformulate the update step (\ref{eq: update step w}) to  
 \begin{equation}\label{eq: update step w and s}
     u_{t+1} = u_{t} +\alpha w_t + s_t
 \end{equation}
 and define the following condition for persistently exciting input with a minimal number of time steps,
\begin{equation} \label{eq: full rank}
        \text{rank}\left(\left[\Delta u_{t-n_u}, \ldots , \Delta u_{t-1}, \alpha w_t + s_t \right] \right) = n_u,
\end{equation}
 where $\Delta u_{t-n_u}, \ldots , \Delta u_{t-1}$ are the $n_u-1$ previous input actuations and $\alpha w_t + s_t$ is the current candidate solution for the next actuation. The idea here is that we first calculate the descent direction and then perturb the corresponding input if the corresponding input is not persistently exciting. When a perturbation is required for (\ref{eq: full rank}) to be satisfied, we want to minimize its effect on the descent direction, measured by its impact on the objective function, by solving:
\begin{subequations}\label{eq: min prob s} 
 \begin{align}
        s_t\left(w\right) = & \underset{s \in \mathbb{R}^{n_u}}{\mathrm{arg\,min}} \quad \frac{1}{2}\left(\nabla_u \Phi \left(u,y\right)s\right)^2 \label{eq: small phi}\\
        \text{s.t.}\quad & \text{rank}\left(\left[\Delta u_{t-n_u}, \ldots , \Delta u_{t-1}, \alpha w_t + s \right] \right) = n_u \label{eq: full rank only s prob}  \\
        \quad & \underline{s} \leq s \leq \overline{s},
 \end{align}
\end{subequations}
where $\underline{s}$ and $\overline{s}$ are upper and lower limits on $s_t$, respectively. With the input actuation reformulation (\ref{eq: update step w and s}), the OFO controller (\ref{eq: orig opt}) becomes 
 \begin{subequations}\label{eq: opt s and w}
      \begin{align}
                w_t =& \underset{w \in \mathbb{R}^{n_u}}{\mathrm{arg\,min}} \lVert w +\nabla_u\Phi^\top\left(u,y\right) \rVert^2_2  \\
                % \text{s.t.} \quad \underline{\boldsymbol{u}} \leq \overbrace{\boldsymbol{u}_t +\alpha\boldsymbol{w}_t}^{u_{t+1}}\leq \overline{\boldsymbol{u}}
                 \text{s.t.}& \quad u_t +\alpha w +s_t\left(w\right) \in \mathcal{U} \\
               % \underline{\boldsymbol{y}}\leq \boldsymbol{y}_t+\alpha \Bar{H}_t\boldsymbol{w}_t \leq \overline{\boldsymbol{y}},
                &\quad y_t+\alpha \nabla_u h_t w+ \nabla_u h_t s_t\left(w\right) \in \mathcal{Y},  %\\
                %&\quad \text{rank}\left(\left[\Delta  u_{t-n_u}, \ldots , \Delta u_{t-1}, \alpha w + s_t \right] \right) = n_u, \label{eq: full rank only s prob}
    \end{align}
 \end{subequations}
where $y_t+\alpha \nabla_u h_t w_t+ \nabla_u h_t s_t\left(w\right)$ is our estimate of the next output and $s_t\left(w\right)$ is the solution of (\ref{eq: min prob s}) given $w_t$. 

In general, the rank function in constraint (\ref{eq: full rank only s prob}) is discontinuous and nonconvex, and general rank-constrained optimization problems are mostly classified as NP-hard \cite{b5}. An iterative rank minimization method is proposed in \cite{b5}, treating the entire matrix as the optimization variable. However, this approach is not suitable for our purposes, as our optimization variables are limited to the last column of the matrix. In \cite{b6}, a convex rank constraint is formulated but is limited to symmetric positive definite matrices, which in general does not hold for the matrix in (\ref{eq: full rank only s prob}). In this paper, we propose a relaxation of constraint (\ref{eq: full rank only s prob}) using linear constraints and penalization. The left nullspace vector of the $n_u-1$ first columns of the matrix in (\ref{eq: full rank only s prob}):
    \begin{equation}\label{eq: v_perp}
        v_\perp = \text{null}\left(\left[\Delta u_{t-n_u}, \ldots , \Delta u_{t-1}\right]^\top \right)
    \end{equation}
provides a vector orthogonal to the span of these columns \cite{strang}. For the new input actuation, $\alpha w_t+ s_t$, to lie outside this span, it must extend in a direction not contained within this span, i.e. the dot product between $\alpha w_t+ s_t$ and $v_\perp$ has to be nonzero:
    \begin{equation}\label{eq: v_perp abs}
        v_\perp^\top\cdot\left(\alpha w_t+ s_t\right)\neq 0.
    \end{equation}
To ensure robustness, this condition is reformulated by introducing a small positive margin parameter, $\epsilon$, resulting in the following inequality:
\begin{equation}\label{eq: abs value}
    \lvert v_\perp^\top \cdot \left(\alpha w_t + s_t\right) \rvert \geq \epsilon.
\end{equation}
To incorporate the nonconvex constraint \eqref{eq: abs value} into a solver designed for linear constraints, we apply a standard lifting technique by introducing two variables $z^+_t$ and $z^-_t$, to represent the absolute value in (\ref{eq: abs value}). For this reformulation to be valid, only one of the variables may be nonzero at a time $t$,  satisfying the following conditions: 
\begin{subequations}\label{eq: z trick}
\begin{align}
    z^+_t + z^-_t  &\geq \epsilon \label{eq: z abs value} \\
    z^+_t &\geq 0 \\
    z^-_t &\geq 0 \\
    z^+_t - z^-_t &= v_\perp^\top\cdot\left(\alpha w_t + s_t\right) \label{eq: z split} \\
    z^+_tz^-_t &= 0. \label{eq: nonlinear z}
\end{align}
\end{subequations}
Constraints \eqref{eq: z abs value} and \eqref{eq: nonlinear z} together ensure that only one of $z_t^+, z_t^-$ is nonzero. To overcome the non-convexity introduced by \eqref{eq: nonlinear z}, we relax \eqref{eq: nonlinear z} by introducing a penalization term $\gamma\left(z^+_t+z^-_t\right)$ into the objective \eqref{eq: small phi}. The relaxation allows formulating \eqref{eq: min prob s} as a convex optimization problem. As demonstrated in Section \ref{sec: example study}, the relaxation provides persistently exciting inputs even for moderate values of $\gamma$.

\subsection{Bilevel Formulation}
To solve \eqref{eq: min prob s} and \eqref{eq: opt s and w}, we propose a bilevel program with (\ref{eq: opt s and w}) as the upper-level problem and (\ref{eq: min prob s}) as the lower-level problem. The rank equality constraint (\ref{eq: full rank only s prob}) is addressed using the reformulation (\ref{eq: z trick}) with the penalty term added to the objective function \eqref{eq: small phi}, relaxing the nonconvex constraint \eqref{eq: nonlinear z}: \newline
\textbf{Upper-level problem:}
\begin{subequations}\label{eq: new opt 3}
\begin{align}
        w_t = &\underset{w \in \mathbb{R}^{n_u}}{\mathrm{arg\,min}} \lVert w + \nabla_u\Phi^\top\left(u,y\right) \rVert^2_2 \\
        % \text{s.t.} \quad \underline{\boldsymbol{u}} \leq \boldsymbol{u}_t +\alpha\boldsymbol{w}_t + \boldsymbol{s}_t \leq \overline{\boldsymbol{u}}
        \text{s.t.} &\quad u_t +\alpha w + s_t \in \mathcal{U} \label{eq: con1-3} \\ 
       % \underline{\boldsymbol{y}}\leq \boldsymbol{y}_t+\alpha \Bar{H}_t\boldsymbol{w}_t  + \Bar{H}_t\boldsymbol{s}_t\leq \overline{\boldsymbol{y}}.
       &\quad y_t+\alpha \nabla_u h_t w  + \nabla_u h_t s_t \in \mathcal{Y}. \label{eq: con2_3}
\end{align}
\end{subequations}
\textbf{Lower-level problem:}
\begin{subequations}
\begin{align}
        \begin{bmatrix}
            s_t \\ 
            z^+_t \\
            z^-_t
        \end{bmatrix} = & \underset{%s\in \mathbb{R}^{n_u}, z^+,z^- \in \mathbb{R}
        \begin{subarray}{c}
s \in \mathbb{R}^{n_u},\\ z^+,z^- \in \mathbb{R}\end{subarray}
        }{\mathrm{arg\,min}}  \frac{1}{2}  \left(\nabla_u \Phi \left(u,y\right)s\right)^2 + \gamma\left(z^++z^-\right)\label{eq: full lower level obj}\\
        \text{s.t.} & \quad \underline{s} \leq s \leq \overline{s}\label{eq: s bounds 3}\\
        & \quad z^+ + z^-   \geq \epsilon \label{eq: full rank first}\\
        & \quad z^+\geq 0\\
        & \quad z^- \geq 0\\
        & \quad z^+ - z^- = v_\perp^\top\cdot\left(\alpha w_t + s\right).\label{eq: full rank last}
\end{align}\label{eq: new opt 32}
\end{subequations}
Bilevel optimization is a class of optimization problems involving two hierarchically nested optimization problems and is in general NP-hard to solve \cite{b8}. However, if the lower-level problem is convex, bilevel optimization problems can be uniquely solved by reformulating the lower-level problem with its KKT-conditions~\cite{b8}. The objective function (\ref{eq: full lower level obj}) is quadratic with a positive semi-definite Hessian, and the constraints \eqref{eq: s bounds 3}–\eqref{eq: full rank last} are linear. As a result, the lower-level problem \eqref{eq: new opt 32} is convex \cite{b11}. However, while the relaxation \eqref{eq: new opt 32} of \eqref{eq: min prob s} does not guarantee the generation of persistently exciting inputs, the perturbations will in practice ensure persistently exciting inputs and in contrast to the controller \eqref{eq: orig opt}, the perturbations are chosen in a minimal disturbing way. The proposed persistently exciting OFO controller (\ref{eq: new opt 3})-(\ref{eq: new opt 32}), with the lower-level problem (\ref{eq: new opt 32}) written using its KKT solution, is:
\begin{subequations} \label{eq: pers. exc. full lag}
\begin{align} 
        w_t =& \quad \underset{
\begin{subarray}{c}
w  \in \mathbb{R}^{n_u}, s \in \mathbb{R}^{n_u}, z^+ \in \mathbb{R}, \\
z^- \in \mathbb{R}, \lambda \in \mathbb{R}^{2n_u+3}, \mu \in \mathbb{R} \end{subarray} }{\mathrm{arg\,min}}\lVert w +  \nabla_u\Phi^\top\left(u,y\right)\rVert^2_2\\
        \text{s.t.}& \quad u_t +\alpha w + s  \in \mathcal{U}\\
       & \quad y_t+\alpha \nabla_u h_tw + \nabla_u h_ts  \in \mathcal{Y}\\
        & \quad \text{A}\begin{bmatrix}
            s^\top & z^+ & z^-
        \end{bmatrix}^\top -\text{b} \leq 0 \\
        & \quad \lambda_i\left(\text{A}\begin{bmatrix}
            s^\top & z^+ & z^-
        \end{bmatrix}^\top -\text{b}\right)_i  = 0, \quad i \in \mathcal{I} \label{eq: complimentarity constraint} \\
        & \quad z^+ - z^- - v_\perp^\top \cdot \left(\alpha w + s \right)  = 0 \label{eq: equal const} \\
    & \quad \nabla_{\{s,z^+,z^-\}} \mathcal{L}\left(w, s, z^+, z^-, \lambda, \mu\right)  = 0,  \label{eq: lag der in full model} 
\end{align}
\end{subequations}
where the Lagrangian, $\mathcal{L}$, is defined as 
\begin{equation}
    \begin{split}
        &\mathcal{L}(w,s,z^+,z^-,\lambda,\mu) = 
        \frac{1}{2}\left(\nabla_u \Phi \left(u,y\right)s\right)^2 \\
        &+ \gamma\left(z^+ + z^-\right) + \lambda^\top\left(A\begin{bmatrix}
            s^\top & z^+ & z^-
        \end{bmatrix}^\top -b\right)  \\
        &+\mu\left( z^+ - z^- - v_\perp^\top \cdot \left(\alpha w + s \right)\right),
    \end{split}
\end{equation}
with $\lambda_i \geq 0$ as the Lagrange multipliers for the inequalities in $\mathcal{I}$, the subscript $i \in \mathcal{I}$ denote the rows in $\text{A}$ and $\text{b}$, and $\mu$ is the Lagrange multiplier for the equality constraint (\ref{eq: equal const}). 

 \begin{figure}[!tbp]
        \centering
        \includegraphics[width=0.50\textwidth, trim=0 0 0 0, clip]{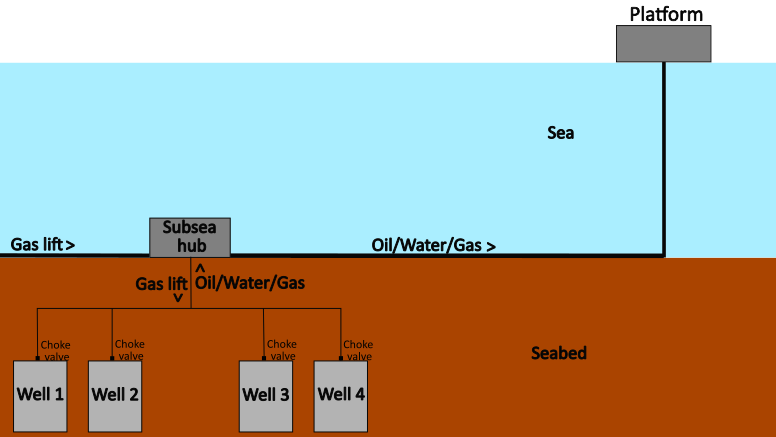}
        \caption{Schematic of the system of oil wells. The subsea hub located on the  seabed distributes incoming gas lift to the oil wells and transports the fluid from the oil wells to the oil platform at sea level.}
        \label{fig: Maria}
\end{figure}
 
\section{Case Study}\label{sec: example study}
The case study is motivated by an oil and gas industrial setup. Figure \ref{fig: Maria} shows a schematic of the system used in this work. Gas lift enters a subsea hub connected to four oil wells, allocating gas lift to each pipeline and sending the produced fluid to the platform at sea level. Gas lift is gas that is injected into the pipeline increasing the production rate by decreasing the density of the fluids. Typically, there is a total limit of available gas lift and there might also be upper and lower limits of injected gas to each well. In the case study, we only consider input constraints, $u \in \mathcal{U}$. The parameters for the case study are given in Table \ref{tab: parameters}.

\subsection{Gas Lift Optimization}
\begin{figure}[!t]
        \centering
        \includegraphics[width=0.50\textwidth, trim=20 0 0 0, clip]{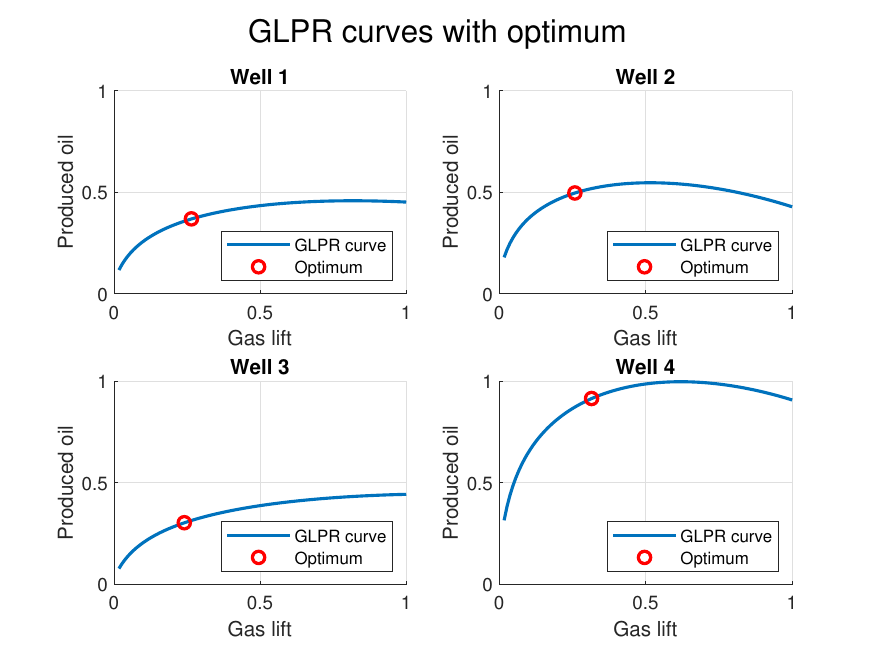}
        \caption{Illustration of the GLPR curves used in the case study. Red circle indicates the optimal operating point for the wells when the total available gas lift is at a certain capacity.}
        \label{fig: glpr}
\end{figure}

Several approaches to real-time optimization of gas lifted oil wells are discussed in \cite{b21}, \cite{b22}, \cite{b23} and \cite{b24}. Oil wells are complex systems that are difficult to model \cite{b23}. Following \cite{b23}, we assume that the wells operate in steady state, so that we can use OFO to allocate the limited amount of gas across the wells. OFO is applicable because it requires only an estimate of the Jacobian of the static input-output mapping. An example of the static input-output mapping is in Fig. \ref{fig: glpr}, named Gas Lift Performance Curves (GLPR) \cite{b19}. Produced oil is on the y-axis for four wells with gas lift along the x-axis. In this paper, the curves are obtained as normalized polynomials. If sufficient gas lift is available, the optimal operating point is where the profit from producing additional oil equals the cost of utilizing more gas lift \cite{b20}. When the amount of gas lift is insufficient for this, the optimum operating points on these curves are where no additional profit can be achieved by redistributing gas lift between wells.
For a system of $N$ oil wells, the outputs from well $i, i = 1,\ldots, N$ are produced oil, $q_\text{o}^i$, gas, $q_\text{g}^i$, and water, $q_\text{w}^i$, $y^i = \begin{bmatrix} q_\text{o}^i & q_\text{w}^i & q_\text{g}^i \end{bmatrix}^\top$, given by a steady-state mapping of the input $y^i = h^i(u^i)$. In this work, it is assumed that the gas and water produced from the well $i$ is a constant fraction of the oil produced from the same well, $r_g^i$ and $r_w^i$, respectively. The inputs to well $i$ are gas lift, $q_{\text{inj}}^i$ and the choke valve opening $v^i$, $u^i = \begin{bmatrix} q_{\text{inj}}^i & v^i \end{bmatrix}^\top$. The choke valve opening ranges from fully open to closed and is typically located where the pipeline is connected to the oil well and affects the production rate by controlling the pressure in the inlet of the pipeline \cite{b19}. The profit from the system of oil wells, $\Phi\left(u,y\right)$, is determined by the oil price, $p_\text{o}$, gas price, $p_\text{g}$, cost of processing water, $-p_\text{w}$ and cost of gas lift, $-p_{\text{inj}}$. Gas lift allocation written in the form of (\ref{eq: mod 0}):
\begin{subequations}\allowdisplaybreaks\label{eq: model}
\begin{align}
    \min_{u, y} \quad  - \sum_{i=1}^{N} &\begin{bmatrix}
        p_{\text{o}} & p_\text{g} & -p_\text{w} 
    \end{bmatrix} y^i  
    + \begin{bmatrix}
        -p_\text{inj} & 0 
    \end{bmatrix} u^i \label{eq: objective function math} \\
    \textrm{s.t. } \sum_{i=1}^{N}{q^i_{\text{inj}}} &\leq \overline{q_{\text{inj}}} \label{eq: model u first}\\
    \underline{q_{\text{inj}}}^i &\leq q_{\text{inj}}^i \leq \overline{q_{\text{inj}}}^i, \quad i = 1, \ldots, N \\
    v^i &\in [0,1], \quad i = 1, \ldots, N \label{eq: model u last}\\
    y^i & = h^i(u^i), \quad i = 1, \ldots, N \label{eq: y},
\end{align}
\end{subequations}
where \eqref{eq: objective function math} denotes the negative profit, $-\Phi\left(u,y\right)$, and constraints (\ref{eq: model u first})-(\ref{eq: model u last}) denote $u \in \mathcal{U}$. The upper limit in \eqref{eq: model u first} is total availability of gas lift and the upper and lower limits in \eqref{eq: model u last} are for each well. The input-output mapping (\ref{eq: y}) is assumed in this work to be a static and unknown function, from which samples $y^i, i = 1 \ldots N$ can be obtained at a given time $t$ and the input and output vectors are stacked in vectors $u = \begin{bmatrix}
    u_1^\top, u_2^\top, \ldots, u_N^\top
\end{bmatrix}^\top$ and $y = \begin{bmatrix}
    y_1^\top, y_2^\top, \ldots, y_N^\top
\end{bmatrix}^\top$.

\subsection{Setup}
For comparison of our proposed persistently exciting OFO controller (\ref{eq: new opt 3})-(\ref{eq: new opt 32}), here called PE OFO, we use the OFO controller (\ref{eq: orig opt}) with zero mean Gaussian noise as perturbation, here called OFO. The controllers have been compared with the same stepsize, $\alpha$ and initialization of the optimization problem and the gradient estimation. Our PE OFO controller is solved using the KKT formulation from (\ref{eq: pers. exc. full lag}). As the OFO controller (\ref{eq: orig opt}) involving Gaussian distribution yields slightly different results for each simulation, the result presented from this method is the mean of 1000 simulations with the standard deviation of the simulations indicated. The simulations are run on a Dell XPS 15 9530 running Windows 11 and
MATLAB R2023b. 

The complementarity constraint \eqref{eq: complimentarity constraint} is known to cause difficulties to nonlinear program solvers, because the Linear Independence Constraint Qualification (LICQ) conditions does not hold \cite{schwartz2011mathematical}. An overview of some approaches developed to solve optimization problems with complementarity constraints may be found in \cite{schwartz2011mathematical}. In this work we used \texttt{fmincon} with the interior point algorithm to solve \eqref{eq: pers. exc. full lag} because of its flexibility and ability to handle infeasibility of the initial guess. While interior point algorithms have been designed to handle complex nonlinear problems \cite{b11}, the analysis of their numerical robustness for problems with complementarity constraints is out of scope of this work.

\subsection{Results}
\begin{figure}[!tbp]
        \centering
        \includegraphics[width=0.47\textwidth, trim=10 3 0 0, clip]{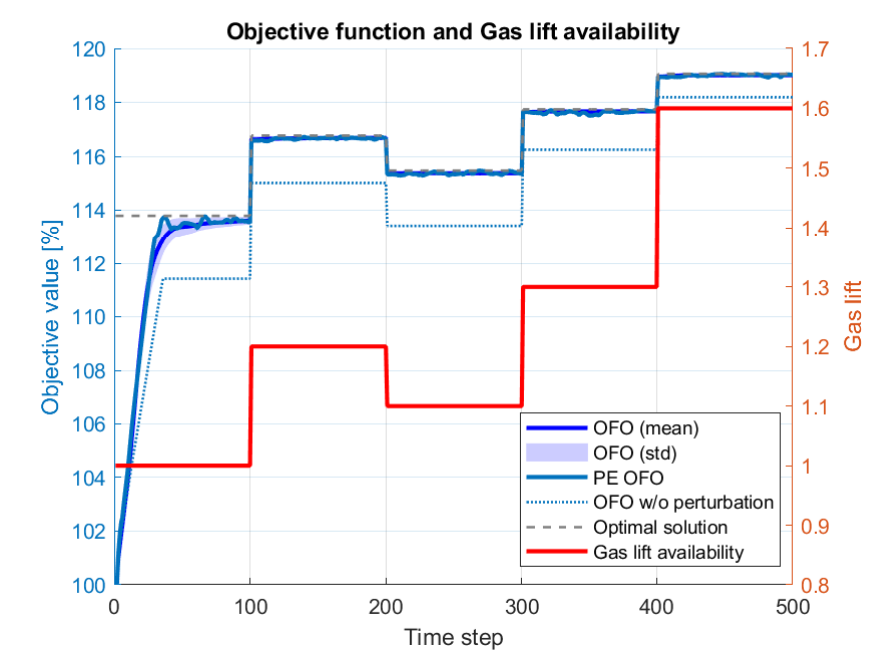}
        \caption{Objective function as a function of time step at the left y-axis, total gas lift availability at the right y-axis in red. Mean value of the OFO simulations is in blue with $\pm$ one standard deviations (lighter blue). The results from the PE OFO controller are in turquoise and the optimal solution is plotted with dotted gray lines. The results from the OFO controller without perturbations is plotted in dotted turquoise. The total gas lift availability is changed as a step every 100\textsuperscript{th} time step.}
        \label{fig: phi stepwise}
\end{figure}

We present the result from a simulation with $N=4$ oil wells with a time-varying availability of gas lift (\ref{eq: model u first}). As we simulate a steady-state map, the time steps are not associated with a given time, but in a physical system, the time steps might be in the range of hours allowing the oil wells to reach steady state. In Fig. \ref{fig: phi stepwise} the objective value at each time step is plotted along the left y-axis for both controllers together with the optimal value given the current availability of gas lift. A simulation of the controller \eqref{eq: orig opt} without perturbation, that is $s_t = 0$ for all $t$, here called OFO without (w/o) perturbation, is also plotted. Along the right y-axis the total gas lift availability at each time step is plotted. The optimal value is from an OFO simulation where the sensitivity of the GLPR curves are known. We observe that both controllers converge to the time-varying optimum, and initially, our proposed PE OFO controller converges faster yielding a 0.01$\%$ higher profit during the simulation. The OFO controller without perturbations does not reach the time-varying optimum and has 1.4$\%$ lower profit during the simulation than the PE OFO controller. Without perturbations, once the total gas lift availability constraint is reached, no new data is sampled, causing the controller to converge sub-optimally due to inaccurate sensitivity estimation. This shows the importance of persistently exciting data to ensure accurate sensitivity estimation to enable optimal convergence. The persistently exciting controllers show close to equal performance in converging and adapting as the total availability of the gas lift changes, meaning that both methods estimate the true sensitivity well. However, for the initial convergence, we observe that the PE OFO controller which perturbs the input in a direction that minimally disturbs the descent direction achieve the fastest convergence, and the OFO controller that perturbs the input with Gaussian noise have a large standard deviation before learning the sensitivity. This large standard deviation is due to the Gaussian perturbations disturbing the input differently at each simulation, with some perturbations leading to notably slower convergence than others. For the first 100 time steps, the PE OFO controller yields some input actuations which moves the solution away from the optimum, indicating that more data is needed to estimate the true sensitivity around the optimum well. For this simulation, our PE OFO controller have no violations of the full rank constraint \eqref{eq: full rank only s prob}, showing that the relaxation in \eqref{eq: new opt 32} works in practice.

\begin{figure}[!tbp]
        \centering
        \includegraphics[width=0.50\textwidth, trim=30 20 25 0, clip]{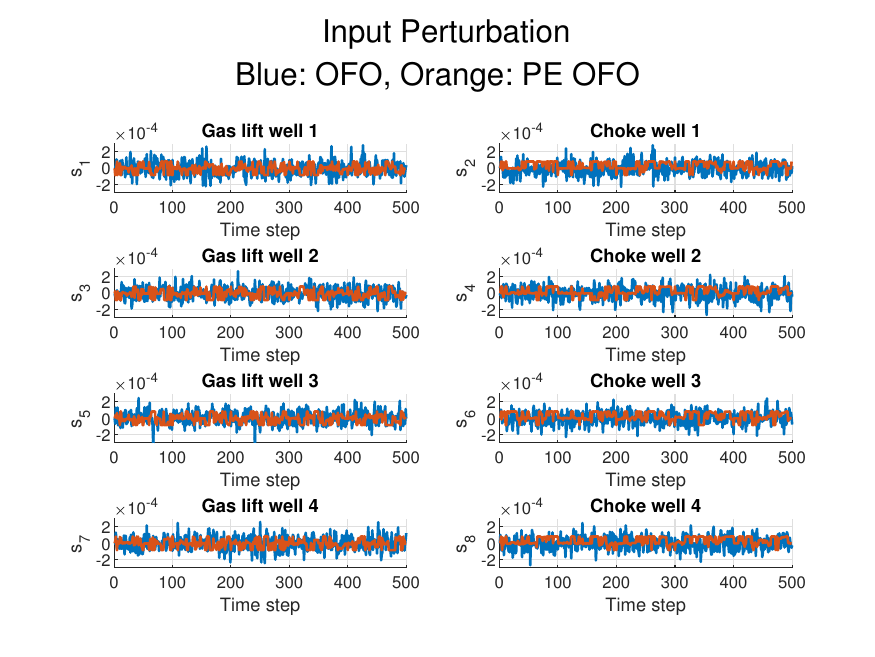}
        \caption{Applied perturbation to each element in the input vector. The standard deviation of the Gaussian distribution is tuned so that the OFO controller reaches the time-varying optimum and divided with the stepsize for comparison.}
        \label{fig: noise2}
\end{figure}

Figure \ref{fig: noise2} shows the perturbation applied to the input in the OFO and the PE OFO controllers. The OFO controller applies perturbation noise at every time step. In contrast, the PE OFO controller applies perturbation only when necessary to ensure persistency of excitation, similar to \cite{b7}. The PE OFO controller perturbs the gas lift inputs more than the choke valve opening inputs. This is due to the choke valves quickly converging to fully open as this increases production and comes without a cost in \eqref{eq: objective function math}. Thus, the choke valve perturbations will be minimal as the descent direction will be perpendicular to the constraints. For the gas lift, each individual gas lift is not at a constraint, but the sum of utilized gas lift is at the constraint. Therefore, the controller perturbs the individual gas lifts more to explore different ways to share available gas lift between wells for optimality. When a perturbation is needed for one input, other inputs are perturbed as well to compensate for the deviation of the descent direction to minimize \eqref{eq: small phi}.

\section{Conclusion}\label{sec: Conclusion}
In this paper we present a persistently exciting input generating Online Feedback Optimization (OFO) controller that perturb the input in a minimal disturbing manner. Our proposed persistently exciting input generating OFO controller has a similar performance to a more traditional OFO controller while we have eliminated the need for random perturbations in a feedback loop. Minimal deviations from the descent direction could also lead to faster convergence, which is especially useful in scenarios where the optimal operating point is time-varying. 

In a physical system, noise would be present, which could contribute to having persistently exciting input-output samples without designing a controller to do so. However, depending on the signal-noise ratio in the system, and considering that not all operating ranges may not be sufficiently explored, our proposed method could be useful in physical systems with disturbances to ensure that all possible directions in the input space are sufficiently explored. Future work could include verifying the method on a physical system with disturbances.

\begin{table}[!tbp]
\caption{Parameters used for the gas lift optimization case study}\label{tab: parameters}
\centering
\setlength{\tabcolsep}{1pt} % Default is 6pt; adjust as needed
\begin{tabular}{cc}
\hline
\textbf{Parameter} & \textbf{Value} \\ \hline
$\underline{q_{\text{inj}}^i}, \underline{v^i}, i = 1, \ldots N$      & 0      \\ 
$\overline{q_{\text{inj}}}^i, \overline{v}^i, i = 1, \ldots N$      & 1      \\ 
$\underline{s}, \overline{s}$      & -0.005, 0.005    \\
$\alpha, \sigma, \epsilon, \gamma$      & 0.001, 5, 10$^{-9}$, 4     \\ 
$\Sigma_{m1},\Sigma_{m2},\Sigma_{m3}$      & $0.01I_{n_u}$     \\ 
$\Sigma_{p1},\Sigma_{p2}$      & $I_{n_u}$     \\ 
$p_o , p_g, p_w, p_{inj}$      & 5, 1, 0.3, 0.7     \\ 
$\text{GLPR}^1$      & $(5\text{log}_{10}(1+25q_{\text{inj}}^1)-0.05(q_{\text{inj}}^1)^2+2v^1)/34.5$    \\ 
$\text{GLPR}^2$      & $(6\text{log}_{10}(1+35q_{\text{inj}}^2)-0.15(q_{\text{inj}}^2) ^2+3v^2)/34.5$    \\ 
$\text{GLPR}^3$      & $(5\text{log}_{10}(1+20q_{\text{inj}}^3)-0.025(q_{\text{inj}}^3) ^2+v^3)/34.5$    \\ 
$\text{GLPR}^4$      & $(10\text{log}_{10}(1+40q_{\text{inj}}^4)-0.175(q_{\text{inj}}^4) ^2+5v^4)/34.5$    \\ 
$r_g^1, r_g^2, r_g^3 ,r_g^4 $      & $0.1, 0.12, 0.1, 0.1$    \\
$r_w^1, r_w^2, r_w^3 ,r_w^4 $      & $0.3, 0.12, 0.2, 0.2$    \\   \hline
\end{tabular}
\end{table}

\end{document}